\documentclass{pasj01}
\draft 
\Received{$\langle$reception date$\rangle$}
\Accepted{$\langle$acception date$\rangle$}
\Published{$\langle$publication date$\rangle$}
\begin{document}

\title{Suzaku observation of Jupiter's X-rays around solar maximum}
\author{Masaki \textsc{Numazawa},\altaffilmark{1,}$^{*}$
Yuichiro \textsc{Ezoe},\altaffilmark{1}
Takaya \textsc{Ohashi},\altaffilmark{1}
Kumi \textsc{Ishikawa},\altaffilmark{2}
Yoshizumi \textsc{Miyoshi},\altaffilmark{3}
Tomoki \textsc{Kimura},\altaffilmark{4}
Yasunobu \textsc{Uchiyama},\altaffilmark{5} 
Daikou \textsc{Shiota}, \altaffilmark{3,6} and
Graziella \textsc{Branduardi-Raymont}\altaffilmark{7}}

\altaffiltext{1}{Department of Physics, 
Tokyo Metropolitan University, 
1-1 Minami-Osawa, Hachioji, Tokyo 192-0397, Japan }

\altaffiltext{2}{The Institute of Space and Astronautical Science (ISAS), 
Japan Aerospace and Exploration Agency (JAXA), 
3-1-1 Yoshinodai, Chuo-ku, Sagamihara 229-8510, Japan}

\altaffiltext{3}{Institute for Space-Earth Environmental Research,
Nagoya University, 
Furo-cho, Chikusa-ku, Nagoya 464-8601, Japan}

\altaffiltext{4}{Department of Geophysics, 
Tohoku University, 
6-3 Aramaki, Aoba-ku, Sendai, Miyagi 980-8578, Japan}

\altaffiltext{5}{Department of Physics, 
Rikkyo University, 
3-34-1 Nishi-Ikebukuro, Toshima-ku, Tokyo 171-8501, Japan}

\altaffiltext{6}{Applied Electromagnetic Research Institute,
National Institute of Information and Communications Technology,
4-2-1 Nukui-Kita, Koganei, Tokyo 184-8795, Japan}

\altaffiltext{7}{Mullard Space Science Laboratory,
University College London,
Holmbury St Mary,
Dorking RH5 6NT, Surrey, United Kingdom}

\email{masaki@phys.se.tmu.ac.jp}

\KeyWords{planets and satellites: individual(Jupiter) ---
X-rays: general}

\maketitle

\begin{abstract}
We report on results of imaging and spectral studies of X-ray emission from Jupiter observed by Suzaku. 
In 2006 Suzaku had found diffuse X-ray emission in 1--5 keV associated with Jovian inner radiation belts. 
It has been suggested that 
the emission is caused by the inverse-Compton scattering by ultra-relativistic electrons ($\sim 50$ MeV) in Jupiter's magnetosphere.
To confirm the existence of this emission and to understand its relation to the solar activity, 
we conducted an additional Suzaku observation in 2014 around the maximum of the 24th solar cycle.
As a result,
we successfully found again the diffuse emission around Jupiter in 1--5 keV and also point-like emission in 0.4--1 keV\@. 
The luminosity of the point-like emission
which was probably composed of solar X-ray scattering, charge exchange, or auroral bremsstrahlung emission
increased by a factor of $\sim 5$ with respect to 2006,
most likely due to an increase of the solar activity. 
The diffuse emission spectrum in the 1--5 keV band was well-fitted with a flat power-law function ($\Gamma = 1.4 \pm 0.1$) as in the past observation,
which supported the inverse-Compton scattering hypothesis.
However, its spatial distribution changed
from $\sim 12 \times 4$ Jovian radius (Rj)
to $\sim 20 \times 7$ Rj.
The luminosity of the diffuse emission  increased by a smaller factor of $\sim 3$. 
This indicates that 
the diffuse emission is not simply responding to the solar activity, 
which is also known to cause little effect on the distribution of high-energy electrons around Jupiter.
Further sensitive study of the spatial and spectral distributions of the diffuse  hard X-ray emission is important
to understand how high-energy particles are accelerated in Jupiter's magnetosphere.
\end{abstract}

%%%%%%%%%%%%%%%%%%%%%%%%%%%%%%%%%%%%%%%%%%%%%%%%%%%%%%%%%%%%%%%%%%%    INTRODUCTION %%%%%%%%%%%%%%
\section{Introduction} \label{sec:intro}
X-ray observatories have discovered X-ray emission from several objects in our solar system \citep{Bhardwaj(07)}.
A study about objects like planets, moons and comets was made feasible for the first time
thanks to an advance of performances of the observatories.
The X-ray observations of the solar system objects showed their active features and their relation to the solar activity, which will
consequently lead us to understand universal relation between stars and planets,
by combining with knowledges from other remote or in-situ measurements about them.
%
%From the past researches,
It has been established that 
there are three types of the mechanisms of X-ray emission from the objects in the solar system:
solar X-ray scattering on the surface or in the atmosphere of the objects, 
charge exchange (CX) reactions caused by collisions between solar wind containing multivalent ions and a planetary atmosphere containing neutral atoms  \citep{Cravens(03)},
and 
non-thermal emission caused by the acceleration of high energy particles.

Jupiter is the largest and the most luminous planet in our solar system in the X-ray band \citep{Metzger(83), Waite(97)}.
Recent X-ray observatories such as Chandra, XMM-Newton, and Suzaku have detected X-ray emission from Jupiter.
As results of the Chandra and XMM-Newton observations,
\citet{Gladstone(02)}, \citet{B-R(04)}, \citet{Elsner(05)}, \citet{B-R(07a)}, and \citet{B-R(07b)} described that 
the solar X-ray scattering occurred dominantly on Jupiter's disk, 
whereas, in Jupiter's auroral region, 
CX emission was caused by heavy ions originated in the solar wind 
or Jupiter's magnetosphere such as Io's plasma source, 
and non-thermal bremsstrahlung emission was caused by energetic electrons in Jupiter's magnetosphere.
The auroral emission is known to be somewhat solar-driven.
Although the majority of the emitting ions/electrons are provided from Jupiter's moons in the magnetosphere, 
some originate from the solar wind. 
The solar wind plasma is strongly accelerated in the Jovian magnetosphere (see e.g. \citet{Bunce(04)}).
%the other is injected by the solar winds.
%
%Moreover, they need to be magnetically accelerated in the ``fast flow'' case of the solar winds, 
%appropriate to high density and intense interplanetary magnetic field,}
%as described 

\citet{Ezoe(10)} have discovered diffuse hard X-ray emission, from Suzaku observation in 2006, around Jupiter 
with a size of $12 \times 4$ Jovian radius (Rj) spatially associated with its radiation belts and Io plasma torus.
It was world-first and had not been observed by Chandra and  XMM-Newton, 
since Suzaku has a comparative advantage in its sensitivity for such extended emission.
They hypothetically considered that
the emission was caused by the inverse-Compton scattering 
by which high energetic electrons ($\sim 50$ MeV) in Jupiter's magnetosphere energized solar visible photons into the X-ray band.
%
%This could clearly explain that no Si K$_\alpha$ line in 2.3 keV existed in its spectrum.
%
However, 
they also mentioned that there was a factor of 7--50 discrepancy
between the $\sim 50$ MeV electron number density required to explain the X-ray intensity of the diffuse emission
and that estimated from the Divine-Garrett empirical charged particle model \citep{D&G(83)} based on the data taken with Pioneer and Voyager.
Also, so far there has been no knowledge about a relation between the detection of the diffuse emission and the solar activity.

The Jovian diffuse hard X-ray emission has several remaining issues. %, as mentioned in \cite{Ezoe(10)} .
Hence, to confirm about a reproducibility of the detection of the
diffuse emission,
we conducted an additional Suzaku observation of Jupiter in 2014 
when the solar activity reached around its maximum of the 24th solar cycle.
In this paper,
we report the results and the relation to the solar activity 
by comparing with the results of the past observation in 2006 when the solar activity was lowering into its minimum.

%%%%%%%%%%%%%%%%%%%%%%%%%%%%%%%%%%%%%%%%%%%%%%%%%%%%%%%%%%%%%%%%%%%    OBSERVATION %%%%%%%%%%%%%%%
\section{Observations}
Suzaku \citep{Mitsuda(07)} observed Jupiter on 15--21 April 2014 with the X-ray Imaging Spectrometer (XIS: \citet{Koyama(07)}).
The XIS consists of two front-illuminated (FI) CCDs of XIS0 and XIS3, and one back-illuminated (BI) CCD of XIS1.
The particle background of XIS is 2--5 times lower by comparing with the instruments onboard Chandra and XMM-Newton 
as shown in figure 5 of \citet{Mitsuda(07)} 
which summarizes the background normalized by the effective area and the field of view.
This indicates that XIS has good sensitivity determined by background for spatially extended emission. 
%
%This is why we utilize only the XIS data in this report in the basically same way as described by \citet{Ezoe(10)}.

Jupiter's motion on the sky ($\sim 20$ arcsec/h in April 2014) required fourteen pointing trims of the spacecraft attitude
during the long observations 
so that the target would not move out of the central region of CCD\@.
The XIS was operated in the normal mode. 
The data reduction was performed on the nominal final version 3.0.22.44 screened data provided by the Suzaku processing facility
by using the HEAsoft analysis package version 6.22.1. 
The total exposure of each XIS FI and BI chip was $\sim$160 ks. 
For spectral fits, response matrices and auxiliary files were generated with \texttt{xisrmfgen} and \texttt{xissimarfgen}.
We treated the optical loading as a negligible problem (see section 2 in \cite{Ezoe(10)}).

The solar activity was going toward its maximum during the observations.
In contrast, 
we were at low activity period just before solar minimum during the past Suzaku observations on 24--28 February 2006.
Therefore, 
we are able to look into the relation between the solar activity and Jupiter's X-rays by comparing these two observations.
We also note that we assumed a distance from Earth to Jupiter as 5.4 AU in the 2014 observations, 
whereas, the past one had been 5.0 AU\@.
Jupiter rose/set after and before Sun in 2014 and 2006, respectively.
The Sun--Jupiter--Earth angle in 2014 was similar to the past observations, $\sim 10^{\circ}$.
We need to subtract diffuse background radiations, 
which are the Soft X-ray Diffuse Background (SXDB) and the Cosmic X-ray Background.
The SXDB consists of Galactic emission from Local Hot Bubble and Milky Way Halo, 
and CXB is the superposition of point sources from distant Active Galactic Nuclei with a rather hard spectrum.
They contaminate Jupiter's X-ray data due to the wide point spread function of Suzaku.
By employing events accumulated in outer regions of the XIS field
as the background,
the above diffuse radiations were subtracted in the spectral data.
We, however, decided to omit
events below 0.4 keV in the following analyses for the 2014 observations,
because Jupiter passed a region where the SXDB is relatively high according to the ROSAT All Sky Survey.

%%%%%%%%%%%%%%%%%%%%%%%%%%%%%%%%%%%%%%%%%%%%%%%%%%%%%%%%%%%%%%%%%%%    RESULTS  %%%%%%%%%%%%%%%%%%
\section{Results}
%%%%%%%%%%%%%%%%%%%%%%%%%%%%%%%%%%%%%%%%%%%%%%%%%%%%%%%%%%%%%%%%%%%    IMAGING   %%%%%%%%
\subsection{Imaging analysis}
\begin{figure*}[H]
%\centering
\includegraphics[width=\textwidth]{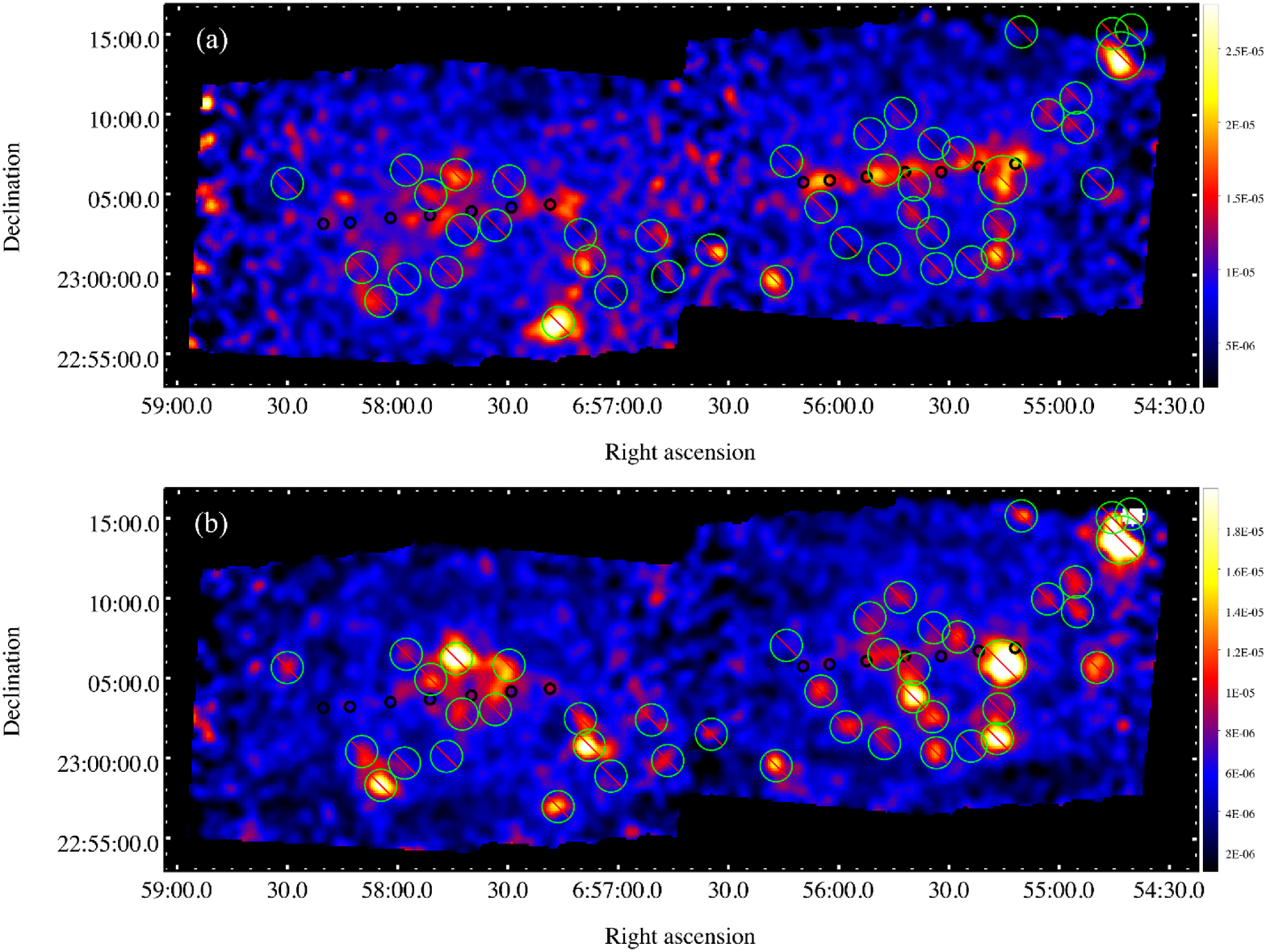}
\caption{Suzaku XIS mosaic images of the vicinity of Jupiter in the (a) 0.4--1 keV (BI) and (b) 1--5 keV (FI) bands, displayed on the J2000.0 coordinates. 
Exposures are corrected and the count unit is counts s$^{-1}$ binned pixel$^{-1}$. 
For clarity, the images are binned by a factor of 8 and smoothed by a Gaussian of $\sigma$ $=$ 5 pixels. 
Black circles show the size and pass of Jupiter during the observations. 
%Considering known pointing uncertainty of Suzaku (Uchiyama et al. 2008), 
%positions of the circles are slightly shifted by $+$25 and $-$25 arcsec in the right ascension (R.A.) and declination (decl.) directions, respectively, 
%in order that the extended emission coincides with Jupiter's footprint. 
Green circles with a red slash mark omitted point source regions when we create the images in Jupiter reference frame coordinates and the spectra (see the text).
}
\label{fig:mosaic}
\end{figure*}

Since Jupiter was moving in the field of view, 
we had to correct for the motion and convert the data coordinates into Jupiter's reference system to detect emission from Jupiter.
We used Jupiter's ephemeris obtained from the Jet Propulsion Laboratory (JPL).
At first, 
before the above-mentioned procedure, 
we examined X-ray images without the attitude correction.
Figure \ref{fig:mosaic} shows mosaic images of BI and FI CCDs in two bands of 0.4--1 and 1--5 keV, respectively.
The images were divided by exposure maps which were generated by the exposure map generator \texttt{xisexpmapgen}.
Both of the images, especially the 0.4--1 keV one, show faint trails 
along Jupiter's motion as an evidence of a detection of Jupiter's X-rays.

There are signatures of emission from point sources in the background beside Jupiter's trail.
In particular, it is noticeable in the 1--5 keV image.
Therefore, we utilized the 1--5 keV image,
checked it against the celestial catalogue of XMM-Newton Serendipitous Source Catalog (3XMM-DR8 version), 
used \texttt{wavdetect} which is a point search program in the CIAO package,
and identified a total of forty four point sources.
The obtained point source positions are shown in figure \ref{fig:mosaic} as green circles with a red slash.
We excluded the circular regions centered at individual sources.
The diameter is equal to the beam size or the half power diameter (HPD) of XIS ($\sim 2$ arcmin).
For safety, we decided to remove X-ray photons from these forty four point sources in both analyses of the 0.4--1 and 1--5 keV images.

We corrected the two band images for Jupiter's motion and created Jupiter's reference frame images. 
For both images in the 0.4--1 and 1--5 keV bands, 
%to take into account the excluded regions, 
we created exposure maps for every 2048 s interval,
corresponding to 256 XIS default time bins (8 s each).
Jupiter's migration length in the 2048 s interval was $\sim 10$ arcsec as estimated from Jupiter's apparent velocity, causing a very small image blur
compared with the Suzaku's HPD ($\sim 2$ arcmin).
We also excluded the forty four point source regions from every exposure map. 
We subsequently corrected the coordinates of all the chopped exposure maps by considering Jupiter's motion 
and reconstructed them into one exposure map for every CCD.
We finally divided the X-ray image by the reconstructed exposure map in Jupiter's reference frame.

\begin{figure*}[H]
%\centering
\includegraphics[width=\textwidth]{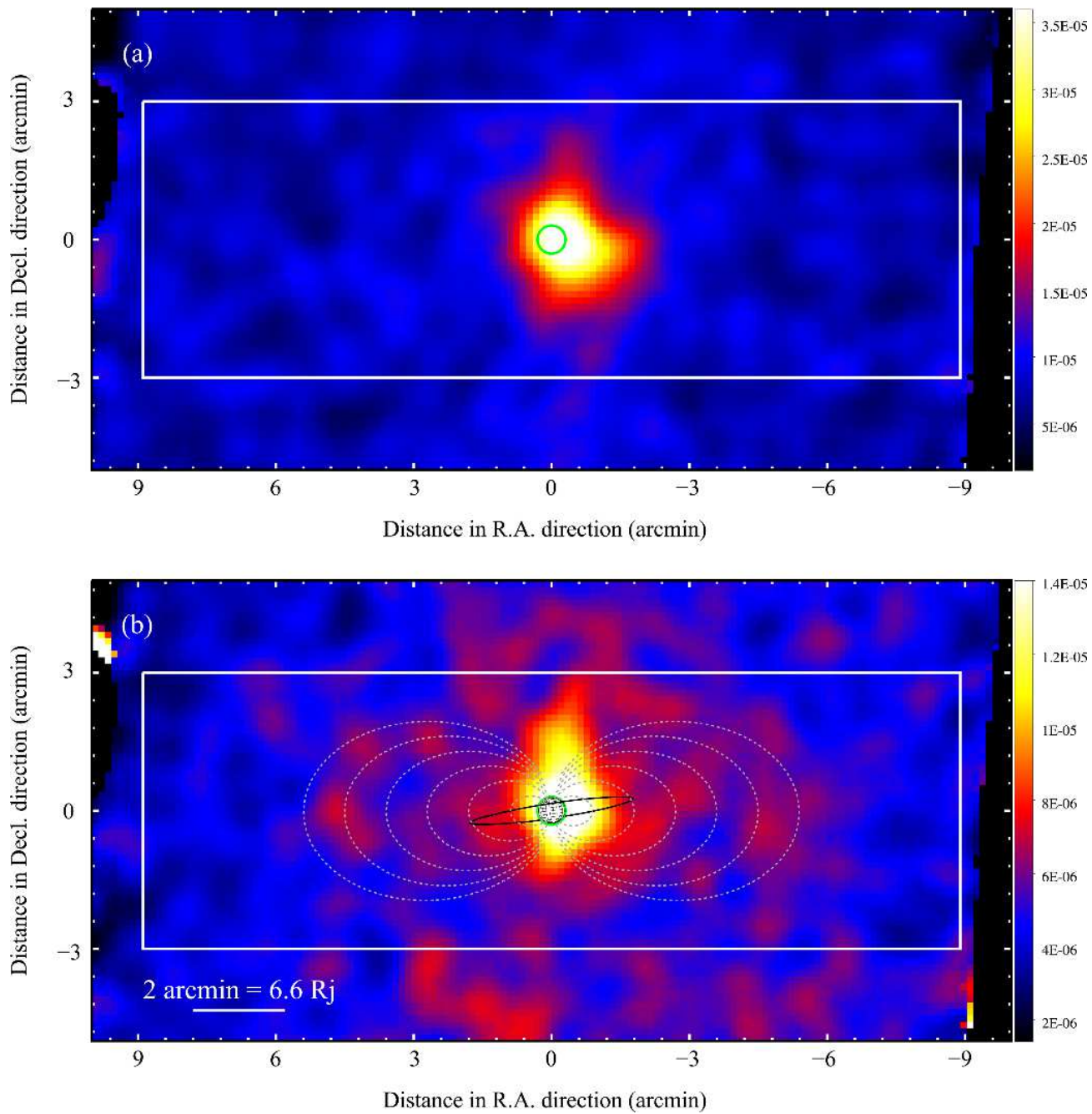}
\caption{Suzaku XIS images after correcting for Jupiter's ephemeris in the (a) 0.4--1 keV (BI) and (b) 1--5 keV (FI) bands. 
The images are binned and smoothed in the same way as figure \ref{fig:mosaic}. 
In panel (a) and (b), a green circle indicates the expected position and size of Jupiter with Jovian radius of18.3 arcsec. 
In panel (b), gray dashed-lines indicate the magnetic field lines at 3, 6, 9, 12, 15, and 18 Jovian radius (Rj). 
A black line is the path traced by Io. 
A schematic diagram of Jupiter was generated by Jupiter Viewer Tool, PDS Rings Node.
%An arrow indicates the direction of the Sun. 
White boxes show the regions used to obtain projection profiles in figure \ref{fig:proj}.
%(A color version of this figure is available in the online journal.)
}
\label{fig:static}
\end{figure*}

\begin{figure*}[H]
 \begin{center}
  \includegraphics[width=\textwidth]{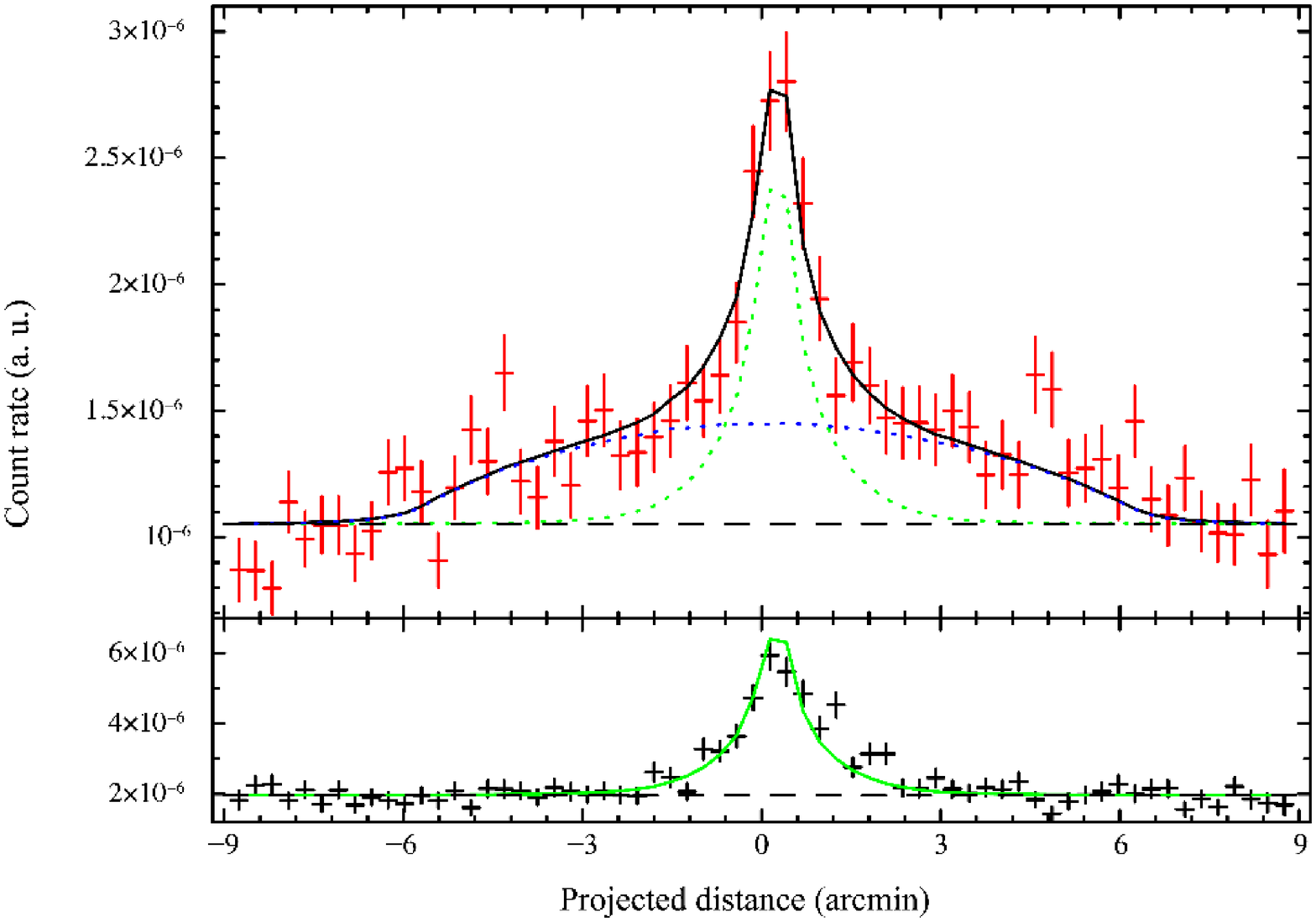}
 \end{center}
 \caption{Projection profiles along the horizontal axis, extracted from box regions in figure \ref{fig:static}.
 Black and red crosses show the 0.4--1 and 1--5 keV data, respectively. 
 Errors are 1$\sigma$ statistical ones. 
 Green lines indicate a simplified model of X-rays from Jupiter's body (18.3 arcsec radius) in 0.4--1 keV (solid line) and 1--5 keV (dotted line).
 A dotted blue line indicates an expected profile of a uniform emission extended over an elliptical region with semi-axes of 6 and 2 arcmin. 
 A solid black line is the sum of the dotted green and blue lines. 
 Dashed black lines are background levels.
 The normalizations and offsets of these lines are tuned by fitting to the data.
 }
 \label{fig:proj}
\end{figure*}

Figure \ref{fig:static} shows the obtained images.
We detected that hard X-ray emission significantly extended over $\sim$20 $\times$ 7 Rj
which is larger than the result of the 2006 Suzaku observation associated with Jovian radiation belts and Io's orbit (see figure 2 in \citet{Ezoe(10)}). 
%%
%The X-ray luminosity of the extended emission arises 
%from (3.6 $\pm$ 0.4) $\times$ 10$^{15}$ erg s$^{-1}$ and (3.3 $\pm$ 0.5) $\times$ 10$^{15}$ erg s$^{-1}$ in 0.2--1.0 and 1.0--5.0 keV, respectively (tbc). 
%%
%Here and below, unless otherwise noted, the uncertainties are 68\% confidence intervals. 
%%
%We assumed a distance to Jupiter of 5.4 AU on 2014 April 15--21. 
%%
%An estimated contamination of the point sources outside of the excluded regions to the extended emission is $\sim$20\% (tbc).
%Hence, most of the emission is due to the extended component.
%%
Figure \ref{fig:static} shows that 
Jupiter's soft X-rays likely correspond to a point-spread function of Suzaku,
in contrast, the hard X-rays are surely more extended than it.
To examine the distribution of the emission quantitatively, 
we produced projection profiles along the horizontal axis as shown in figure \ref{fig:proj}.
We simulated emission distributions from Jupiter's body (disk and aurorae) and from its surroundings
by presuming their shapes to be a small circle and an ellipse, respectively.
We found that the profile can be roughly represented by a simulated model using \texttt{xissimarfgen}, 
which consists of a uniform circular emission with Jovian radius of 18.3 arcsec
and a uniform elliptical emission with semi-axes of 6 and 2 arcmin (20 and 7 Rj).
The hard emission is surely extended over a wide region ($> 6$ Rj). 
On the other hand, 
the soft emission was marginally consistent with the small circular emission from Jupiter's body. 
%
%There is an excess on the right side of the soft X-ray peak, 
%which suggests emission from the IPT that has been reported by Elsner et al. (2002) and \cite{Ezoe(2010)}.

%%%%%%%%%%%%%%%%%%%%%%%%%%%%%%%%%%%%%%%%%%%%%%%%%%%%%%%%%%%%%%%    SPECTRUM    %%%%%%%%%%%%%%%%%%%%
\subsection{Spectral analysis}
In order to investigate the characteristics of Jupiter's X-ray emission,
we analyzed the spectrum by extracting photons from the extended emission region.
We excluded the above-mentioned forty four point sources in the background from each of the event files at first.
In the same way as in generating the exposure maps,
we partitioned the original event file into several parts of 2048 s duration,
and then 
extracted source photons from a circle with a radius of 6 arcmin and background ones from the surrounding annulus with an outer radius of 8 arcmin.
Both regions are centered on the path of Jupiter at any time.
Consequently, 
the source region includes both  Jupiter's body and diffuse emission
because of the limitation of Suzaku's angular resolution (HPD of $\sim 2$ arcmin) which hampers us to separate them spatially.
In the following spectral fits, we used \texttt{XSPEC} version 12.9.0.
The diffuse background component was subtracted using the background photons in the outer annular region.
The apparent size of Jupiter's body is small against the source region, and we may ignore that Jupiter occults a part of the diffuse background radiation. 

\begin{figure*}[H]
 \begin{center}
  \includegraphics[width=\textwidth]{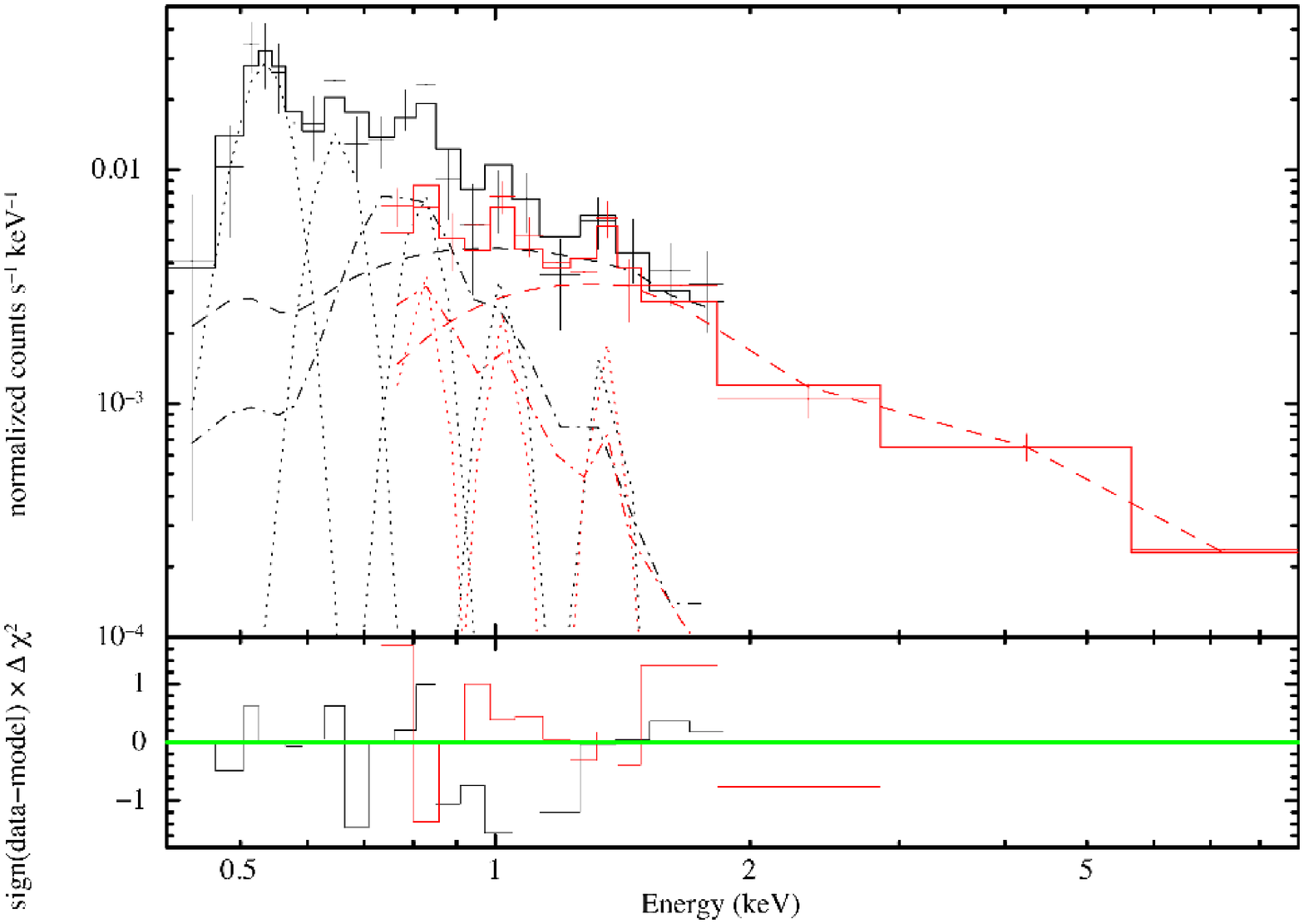}
 \end{center}
 \caption{Background subtracted BI (black) and FI (red) spectra of the extended emission region, 
 compared with the best-fit models summarized in table \ref{tab:spec} (solid lines). 
APEC model, a power-law function, and Gaussians are shown in dash-dotted, dashed, and dotted lines, respectively.
Energies of the Gaussian lines (dotted curves) are 0.53$\pm$0.01, 0.65, 0.82, 1.02, and 1.35 keV, 
representing 
a possible N\,\emissiontype{VII} and O\,\emissiontype{VII} mixture, %\,\emissiontype{VII}, 
O\,\emissiontype{VIII Ly$\alpha$}, 
Fe\,\emissiontype{XVIII}, 
Ne\,\emissiontype{X}, 
and Mg\,\emissiontype{XI} lines, respectively.
The line energies of the latter four lines are fixed in the spectral fits.
 %Purple, green, and blue dashed lines plotted for the FI spectrum are Jupiter’s auroral continuum emission models in XMM-Newton observations (Branduardi-Raymont et al. 2007a).
 }
\label{fig:spec}
\end{figure*}

The spectrum thus obtained is shown in figure \ref{fig:spec}.
It seems to show several lines, with a thermal plasma continuum in the lower energy side and a continuum extending up to $>5$ keV.
We firstly tried to fit a model consisting of five Gaussians combined with an APEC model and a power-law function to the spectra.
We included no absorbing column in the model, as expected from the proximity of the source object. 
The five Gaussians had fixed line energies: 0.57, 0.65, 0.82, 1.02, and 1.35 keV 
which are the known values of 
O\,\emissiontype{VII}, 
O\,\emissiontype{VIII Ly$\alpha$}, 
Fe\,\emissiontype{XVIII}, 
Ne\,\emissiontype{X}, 
and Mg\,\emissiontype{XI} lines, 
respectively.
We assumed that O\,\emissiontype{VII} and O\,\emissiontype{VIII Ly$\alpha$} lines are CX emission from Jupiter's aurorae,
whereas Fe\,\emissiontype{XVIII},  Ne\,\emissiontype{X},  and Mg\,\emissiontype{XI} lines are scattered solar coronal emission from Jupiter's disk.
The first fit to the data gave $\chi^2/\nu$ of $\sim$ 1.12.
However,
we found a difference around 0.5 keV between the model and the spectra.
Therefore, 
we next allowed the line energy of the first gaussian at 0.57 keV to be free 
and fitted again to the same spectrum.
This second fitted model represented the data better than the first one with $\chi^2/\nu$ of $\sim$ 0.70.
We employed this second model as the best-fit model because of the $\chi^2/\nu$ value.
Table \ref{tab:spec} lists the best fit parameters and 90\% confidence errors.
The obtained $\Gamma$ of the power-law function is $1.4 \pm 0.1$,
consistent with the spectra obtained in the 2006 Suzaku observation.
This quite flat power-law is thought to be corresponding to the non-thermal emission discovered from the past Suzaku observation.

%In contrast, 
The line energy of the first Gaussian in the best-fit model is 0.53 keV, which does not correspond to any likely value caused by the CX reaction.
It is supposedly considered that the emission in this region is a line complex of oxygen and nitrogen ions.
Then, we finally tried to fit with six Gaussians by adding one Gaussian, whose line energy was fixed at 0.50 keV to represent the N\,\emissiontype{VII} line, to the first model.
This third model fitted the spectrum as well as the first one with $\chi^2/\nu$ of $\sim 0.92$,
but showed a slight remaining discrepancy around 0.5 keV\@.
%
%Although we were not able to identify what kind of line emissions, 
%we that all of this lines originate from CX reaction between Jupiter's atmosphere and the ions from the solar wind or Io's plasma source.
%

\begin{table*}[H]
\tbl{
Best co-fit parameters (and their 90\% confidence errors) for BI (0.4--2 keV) and FI (0.7--9 keV) spectra of the extended emission region.
}{
%\begin{center}
\begin{tabular}{l c c }\\ \hline \hline
APEC\footnotemark[$*$] 	& kT\footnotemark[$\dagger$] 		& Norm\footnotemark[$\ddagger$] 	\\ %\hline
		& 0.43$_{-0.08}^{+0.09}$  & 19.5 $\pm$ 5.2  	\\ \hline
Power-law	& $\Gamma$ (Photon index)		& Norm\footnotemark[$\ddagger$]		\\ %\hline
		& 1.44$_{-0.11}^{+0.13}$	& 27.2 $\pm$ 2.7			\\ \hline
Gaussian\footnotemark[$\S$] 	& Line energy\footnotemark[$\|$]		& Flux\footnotemark[$\#$]			\\ %\hline
		& 0.53$\pm$0.01	& 76.0 $\pm$ 22.3			\\
		& 0.65		&	21.7 $\pm$ 8.4				\\
		& 0.82 		&	6.6 $\pm$ 2.9				\\
		& 1.02		&	2.4 $\pm$ 1.4				\\ 
		& 1.35 		&	1.2 $\pm$ 0.9				\\\hline
$\chi^2$/d.o.f\footnotemark[$**$]		& 17.58 / 25 & \\ \hline
\end{tabular}
} \label{tab:spec}
\begin{tabnote}
\footnotemark[$*$] APEC abundance parameter is set by \texttt{aspl} \citep{Aspl(09)}.\\
\footnotemark[$\dagger$] APEC temperature in keV.\\
\footnotemark[$\ddagger$] Normalization in units of 10$^{-6}$ ph cm$^{-2}$ s$^{-1}$ keV$^{-1}$.\\
\footnotemark[$\S$] The widths of Gaussians are fixed at 0.01 keV in the fits.\\
\footnotemark[$\|$] Energy of the emission features in keV (fixed in the fits other than the line around 0.53 keV).\\
\footnotemark[$\#$] Total flux in the line in units of 10$^{-6}$ ph cm$^{-2}$ s$^{-1}$.\\
\footnotemark[$**$] $\chi^{2}$ value and degrees of freedom.\\
\end{tabnote}
\end{table*}

%%%%%%%%%%%%%%%%%%%%%%%%%%%%%%%%%%%%%%%%%%%%%%%%%%%%%%%%%%%%%%%%%%%    DISCUSSION %%%%%%%%%%%%%%%%
\section{Discussion: Comparison with the past Suzaku observation in 2006} 
We have detected again the diffuse X-ray emission around Jupiter in the 1--5 keV band,
but also found differences in the results of both imaging and spectral analyses from the past ones.
This may be attributed to a relation between Jupiter's X-ray emission and the solar activity.
In order to study the relation,
we quantitatively compared X-ray luminosities of both observations in 2014 and 2006.
We utilized the best-fit model in the spectral analysis to calculate the luminosities.
We estimated the luminosities using the model components
since the intensities predicted by the model correspond well to the observed net count rates of the spectra within  $1 \sigma$ errors in 0.4--1 and 1--5 keV bands.
We estimated the luminosities for the observation in 2014 as ($14.4 \pm 1.2$) and ($10.4 \pm 0.6) \times 10^{15}$ ergs s$^{-1}$ in 0.2--1 and 1--5 keV, respectively.
%
%\citet{Ezoe(10)} estimated the luminosities for the observation in 2006 
%as ($3.6 \pm 0.4$) and ($3.3 \pm 0.5) \times 10^{15}$ ergs s$^{-1}$ in 0.2--1 and 1--5 keV, respectively.
%%
Since no APEC model had been fitted to the spectrum in the 2006 observations by \citet{Ezoe(10)}, 
we carried out a fit to this spectrum with a new model consisting of an additional APEC component, two Gaussians with line energies fixed at 0.24 and 0.56 keV, and a power-law with $\Gamma$ fixed at 1.4.
The temperature $kT$ of the APEC model was fixed at 0.43 keV (the same value in the 2014 observation), 
to avoid an overlap with other components.
As a result, 
the luminosities for the 2006 observations were estimated from the re-fitted model as ($3.9 \pm 0.6$)  and ($3.1 \pm 0.6) \times 10^{15}$ ergs s$^{-1}$ in 0.2--1 and 1--5 keV, respectively.
%

%\subsection{\textcolor{red}{The soft X-rays:  below 1 keV}}
The soft X-ray luminosity significantly increased overall by a factor of $3.7 \pm 0.6$ from 2006 to 2014.
We investigated a cause of this increase by considering the two regions of the soft X-ray emission, i.e.\ Jupiter's disk and aurorae.
From figures \ref{fig:static} and \ref{fig:proj}, 
we found that 
the soft X-ray image was consistent with point-like emission as small as Jupiter's size (1 Rj $= 18.3$ arcsec),
which implied that they were emitted from Jupiter's body.
The CX lines are not expected to be dominant in Jupiter's disk but rather are produced in the auroral region.
From figure \ref{fig:spec} and table \ref{tab:spec}, we found that 
the solar X-ray scattering on the surface of Jupiter's disk, which was dominant in the soft band, 
was well represented by the APEC model with $kT$ of 0.43 keV in 2014, 
roughly corresponding to the solar coronal temperature.
Based on our new spectral fit to the spectrum in 2006,
%Although no APEC model had been fitted to the spectra in 2006,
the emission with $kT = 0.43$ keV, regarded as Jupiter's disk emission, was then very weak.
%
%The soft disk X-ray luminosity could be estimated by subtracting contributions from the two Gaussians
%%,(3.9 $\pm$ 0.6) and (2.0 $\pm$ 0.5) $\times$ 10$^{-5}$ photons cm $^{-2}$ s$^{-1}$, 
%from the total soft X-ray luminosity in 0.2--1 keV \citep{Ezoe(10)}.
%
%This may give an underestimation of the disk intensity, since the CX line intensities would contain small amount of the continuum emission.
%The soft disk X-ray luminosity only contributed by the APEC model could be estimated and be compared somewhat quantitatively 
%by using the re-fitted model for the 2006 observations.
%
The soft disk X-ray luminosity thus estimated changed 
from ($0.6 \pm 0.4) \times 10^{15}$ ergs s$^{-1}$ in 2006 
to ($3.5 \pm 0.5) \times 10^{15}$ ergs s$^{-1}$ in 2014,
indicating an increase by a factor of $5.6 \pm 3.3$.
The solar 1--8 \AA\ (1.5--12.4 keV) X-ray flux measured by the GOES satellite series also increased
from $\sim 5 \times 10^{-8}$ W m$^{-2}$ in 2006 (GOES 12)
to $\sim 1 \times 10^{-6}$ W m$^{-2}$ in 2014 (GOES 15).
This tendency was consistent with the variation of the Jupiter's disk X-ray power (0.2--2 keV) shown in figure 5 of \citet{B-R(10)}.

%\textcolor{red}{We want to talk about a relation between CX lines and solar winds.}
The luminosity corresponding to the CX emission also varied.
As a result of a simple comparison of the luminosities of the CX emission in 0.4--1 keV,
i.e., between the O\,\emissiontype{VII} intensity in 2006 and the sum of
the $\sim 0.53$ keV and O\,\emissiontype{VIII Ly$\alpha$} components in 2014,
we found that the luminosity significantly increased by a factor of $6.1 \pm 2.1$ 
from ($1.2 \pm 0.4) \times 10^{15}$ ergs s$^{-1}$ in 2006
to ($7.2 \pm 1.0) \times 10^{15}$ ergs s$^{-1}$ in 2014.
The observed increase of the CX emission is expected to be caused by variations of the solar wind parameters, such as density, pressure, and magnetic field. 
There also occurred an event of the Coronal Mass Ejections on April 2, 2014, 
which could have caused a transient increase of Jupiter's CX emission during our observation period in April 15--21.
In fact, an MHD simulation for global heliosphere SUSANOO (\citet{Shiota(14), S&K(16)}) suggested 
that solar wind density and pressure at Jupiter orbit on average increased significantly between these observations. 
%% from $\sim$0.19 cm$^{-3}$ to $\sim$0.34 cm$^{-3}$. 
%
This can qualitatively support our view that the CX emission is an important process for the X-ray emission of Jupiter.
In summary, we consider that the solar activity (variations of solar X-rays and solar wind parameters) has caused the rise of the luminosity of the soft X-rays for both Jupiter's disk and aurorae.
%
%CX contribution 
%
%2006  
%
%2.2944048959205423 $\pm$ 0.3497891295141225
%
%2014
%
%7.184710839014071 $\pm$ 1.0467406217515276
%
%3.131404945913646, 0.6603295306479475

%\subsection{\textcolor{red}{The hard X-rays: 1--5 keV}}
We calculated a diffuse intensity ratio 
defined as the intensity of the simulated ellipse-shape emission (blue dotted curve in figure 3) divided by
that of the small circular emission (green dotted curve), 
by integrating counts in each of these areas in the 1--5 keV projection profiles.
The diffuse intensity ratios in 2014 and 2006 were estimated at 1.8 and 3.2 from figure \ref{fig:proj} and figure 3 in \citet{Ezoe(10)}, respectively.
%
%This implies that the component of Jupite's body (disk$+$aurorae) increases against the past observation in 2006, because of solar activity.
%
These ratios enable us to separate the total hard X-ray luminosity
into two parts; which are Jupiter's body (disk$+$aurorae), represented by the small circle, and the diffuse emission, as the ellipse shape component, respectively.
The hard X-ray luminosity coming from Jupiter's body increased
from ($0.7 \pm 0.1) \times 10^{15}$ ergs s$^{-1}$ in 2006
to ($3.8 \pm 0.2) \times 10^{15}$ ergs s$^{-1}$ in 2014
by a factor of $5.1 \pm 1.1$,
which is in the same sense as the soft X-ray luminosity did.
The emission was contributed for the most part by the solar X-ray scattering on the disk %and both of the CX 
and the bremsstrahlung emission in the aurorae.
Although it was difficult to distinguish the disk from the aurorae component due to the spatial resolution of Suzaku,
the latter had been suggested to be related to the solar activity in terms of the solar wind parameters \citep{Bunce(04), B-R(07a)}.
Therefore, 
considering that the solar X-ray scattering and CX emission in the soft band
have increased significantly,
we conclude that the hard X-ray luminosity of Jupiter's body was also raised by the increase of the solar activity.

The diffuse hard X-ray luminosity, on the other hand, varied
from ($2.4 \pm 0.5) \times 10^{15}$ ergs s$^{-1}$ in  2006
to ($6.6 \pm 0.4) \times 10^{15}$ ergs s$^{-1}$ in 2014
by a factor of $2.8 \pm 0.6$.
The variation amplitude is smaller than that of the soft X-ray luminosity or the hard X-ray luminosity coming from Jupiter's body.
Moreover,
the spatial size of the diffuse X-ray emission in 2006 and 2014 changed from
$12 \times 4$ Rj to $20 \times 7$ Rj by approximating with the ellipse-shape region to reproduce the projected profile.
This suggests that the diffuse hard X-rays may not have a simple relation to the solar activity.
\citet{Ezoe(10)} have suggested the hypothesis that 
the diffuse X-rays were produced by the inverse-Compton scattering of solar visible photons 
by high-energy ($\sim 50$ MeV) electrons in Jupiter's magnetosphere.
It has also been suggested that
the $\sim 50$ MeV electrons were related to high-energy electrons (a few tens MeV) 
responsible for the Jovian synchrotron radio emission (JSR) closer to Jupiter ($<2$ Rj),
although this has a size different from the X-ray emission \citep{Miyoshi(99), Bolton(02)}.
\citet{Bolton(89)}, \citet{S-C(08)} and \citet{Han(18)} indicated that 
a long-term variation of the JSR tends to be out of sync with the solar activity 
after 1994 when Comet Shoemaker Levy 9 impacted on Jupiter,
and the JSR intensity varied by a factor of $<2$ from its minimum to maximum.
However, as \citet{Han(18)} mentioned, 
there is a problem that no sufficient data were gathered after 2005 nor sufficient information obtained about a short-term variation of the JSR\@.
The short-term variation of the JSR (e.g.\ \cite{Miyoshi(99)}) or of the few tens MeV electrons (which is still unclear)
may be a useful clue to understand the variation of the luminosity of the diffuse hard X-ray emission not simply following the solar activity.
Also, 
we have another serious issue that, to account for the hard X-ray luminosity, 
the number density of the high energy electrons in Jupiter's magnetosphere significantly exceeds the level predicted by the current model, 
as pointed out by \citet{Ezoe(10)}.

The uniform elliptical shape was assumed as Jupiter's diffuse component % in this analysis 
to separate it from the emission of Jupiter's body.
The actual distribution of ultra-relativistic particles around Jupiter 
would not be simply uniform, 
but is likely to be denser as one goes nearer to Jupiter \citep{D&G(83)}.
Thus, 
there is a possibility that 
a part of the hard X-ray emission coming from the Jupiter's body estimated in this paper 
is due to the inverse Compton scattering by the dense particles near Jupiter, 
on top of the solar X-ray scattering and the bremsstrahlung emission, as discussed above.
The large change of this emission from 2006 to 2014,
as mentioned earlier,
suggests that the contribution of the inverse Compton scattering would be a small fraction.

In brief, 
we need more detailed information such as spatial and spectral distributions with a moderate resolution about the diffuse X-ray emission (as well as JSR).
\citet{Ezoe(13)} proposed a future astronomical mission named Jupiter X-ray telescope array (JUXTA),
which aims at the first in-situ measurement of Jupiter and its neighborhood with X-ray instruments.
Therefore, we expect the JUXTA and other future missions to bring us more information from Jupiter and its surrounding region.
%
%Although we have the remaining problems unexplained at this time, 
%we considered that the results were possible to corroborate the hypothesis.
%%%%%%%%%%%%%%%%%%%%%%%%%%%%%%%%%%%%%%%%%%%%%%%%%%%%%%%%%%%%%%%%%%%    DISCUSSION %%%%%%%%%%%%%%%%
\section{Summary}
A diffuse hard X-ray emission around Jupiter was discovered by a $\sim 159$ ks Suzaku observation in 2006.
A hypothesis has been suggested that
the inverse-Compton scattering process between solar light and ultra-relativistic ($\sim 50$ MeV) electrons in Jupiter's magnetosphere 
produces the diffuse emission \citep{Ezoe(10)}.
We conducted an additional observation by Suzaku in 2014 with a similar exposure time of $\sim 160$ ks
to investigate the reproducibility of detection of the diffuse hard X-ray emission and a possible dependency of the solar activity on Jupiter's X-rays.
The solar activity reached around its maximum in 2014, whereas it was going toward its minimum in 2006.

As a result, 
we successfully detected again the diffuse X-ray emission in 1--5 keV 
which was spectrally well-fitted with a flat power-law function with $\Gamma$ of $\sim 1.4$, 
supporting the hypothesis of non-thermal emission caused by the inverse-Compton scattering 
of solar photons by high energy electrons with energy $\sim 50$ MeV\@.
We, in addition, separated Jupiter's X-rays into two components contributed by a point-like emission from Jupiter's body and the diffuse emission around Jupiter.
We also compared these results with those of the 2006 observation in terms of luminosity.
The luminosity of Jupiter's body (consisting of Jupiter's disk and aurorae) increased in line with the solar cycle,
whereas
that of the diffuse emission did not change by a similar factor.
The size of the diffuse emission, estimated by fitting with ellipse-shape to the projected profile in 1--5 keV, was found to
change from $12 \times 4$ to $20 \times 7$ Rj between 2006 and 2014.

%We thought that these results are possible to support the hypothesis.
Our results can be interpreted as supporting the hypothesis put forward in \citet{Ezoe(10)}.
This is marginally consistent with the fact that JSR, 
thought to be generated by similar high energy electrons near Jupiter as the diffuse hard X-ray emission,
varies within a factor of 2 from its minimum to maximum
and shows a tendency to be out of sync with the solar activity after 1994.
We really hope that 
future astronomical missions with higher performance will clarify the mechanism  of producing the diffuse emission
and that 
the diffuse emission can  be used as a useful probe to monitor high energy particles in Jupiter's magnetosphere.

\begin{ack}
This research was partially supported by the MEXT Grant-in-Aid No.\ 17J05475.
The authors thank useful comments from the referee, Dr. Thomas Cravens.
\end{ack}
%%%%%%%%%%%%%%%%% 

\end{document}